\documentclass[epsfig,12pt]{article}
\usepackage{epsfig}
\usepackage{graphicx}

\usepackage{latexsym}
\usepackage{amsmath}
\usepackage{amssymb}
\usepackage{relsize}
\usepackage{geometry}
\def\beq{\begin{equation}}
\def\eeq{\end{equation}}
\def\beqn{\begin{eqnarray}}
\def\eeqn{\end{eqnarray}}

\newcommand{\ntwo}{${\mathcal N}=2\,$}

\newcommand{\ce}{{\mathcal E}}

\newcommand{\vp}{\varphi}
\newcommand{\pt}{\partial}

\newcommand{\cde}{{\mathcal D}}

\newcommand{\gsim}{\lower.7ex\hbox{$
\;\stackrel{\textstyle>}{\sim}\;$}}
\newcommand{\lsim}{\lower.7ex\hbox{$
\;\stackrel{\textstyle<}{\sim}\;$}}

\newcommand{\p}{\partial}

\def\slashed#1{\setbox0=\hbox{$#1$}             
   \dimen0=\wd0                                 
   \setbox1=\hbox{/} \dimen1=\wd1               
   \ifdim\dimen0>\dimen1                        
      \rlap{\hbox to \dimen0{\hfil/\hfil}}      
      #1                                        
   \else                                        
      \rlap{\hbox to \dimen1{\hfil$#1$\hfil}}   
      /                                         
   \fi}                                        %

\begin{document}


\begin{titlepage}

\begin{flushright}
FTPI-MINN-13/19, UMN-TH-3208/13, ITEP-TH-16/13\\
\end{flushright}

\vspace{1cm}

\begin{center}
{  \Large \bf  Revisiting the Faddeev-Skyrme Model\\[2mm] and Hopf Solitons}
\end{center}
\vspace{1mm}

\begin{center}

 {\large
 \bf   A.~Gorsky$^{\,a,b,d}$,  M.~Shifman$^{\,b}$ and \bf A.~Yung$^{\,\,b,c}$}

\vspace{3mm}

$^a$
{\it Theory Department, ITEP, Moscow, Russia}\\[1mm]
$^b${\it  William I. Fine Theoretical Physics Institute,
University of Minnesota,
Minneapolis, MN 55455, USA}\\[1mm]
$^{c}${\it Petersburg Nuclear Physics Institute, Gatchina, St. Petersburg
188300, Russia}\\[1mm]
$^{d}${\it Moscow Institute for Physics and Technology, Dolgoprudny, Russia
}
\end{center}

\vspace{1cm}

\begin{center}
{\large\bf Abstract}
\end{center}

We observe that the Faddeev-Skyrme model emerges as a low-energy limit
of scalar QED with two charged scalar fields and a selfinteraction potential
of a special form (inspired by supersymmetric QCD). Then we discuss
possible Hopf solitons of the``twisted-toroidal" type. 

\end{titlepage}

\newpage


\section{Introduction}
\label{intro}
\setcounter{equation}{0}

Many field theories, in particular, supersymmetric Yang--Mills theories, 
support topologically stable solitons. Their stability is due to the existence of
certain topological charges (in case of supersymmetry they are usually related to central
charges of the relevant superalgebra \cite{1}).
In such cases one can perform
 the Bogomol'nyi completion \cite{2} for the energy functional (in the instanton case, for the action)
  which 
selects the filed configuration corresponding to the minimal   energy in the sector with
the given topological charge. Well-known examples are the Abrikosov--Nielsen--Olesen
(ANO) strings \cite{ANO}, whose topological stability is due to 
 $\pi_1({\rm U}(1))=Z$, instantons in the two-dimensional
 CP(1) model \cite{BP} whose topology is determined by 
 $\pi_2({\rm SU}(2)/{\rm U}(1))=Z$, and the
 Belavin-Polyakov-Schwartz-Tyupkin instantons \cite{BPST}
 in four-dimensional Yang--Mills theory
 whose topological classification
 is based on  $\pi_3({\rm SU}(2) )=Z$.
  Faddeev and Niemi discovered \cite{Fadd} a novel class of solitons, of the knot type,
  whose stability is due to the existence of the Hopf topological invariant.
  
The model with the solitonic knots considered by Faddeev and Niemi
is a deformed O(3) nonlinear sigma
model in four dimensions 
\begin{equation}
\mathcal{L} =\frac{F^{2}}{2}\,\, \partial_{\mu}\vec{S}
\,\partial^{\mu}\vec{S} -\frac{\lambda}{4}\,
\left(\partial_{\mu} \vec{S}\times\partial_{\nu}\vec{S} \right)
\cdot\left(\partial^{\mu}\vec{S} \times\partial^{\nu}\vec{S} \right)\,.
\label{senza}
\end{equation}
where the three-component field $\vec S$ is an ``isotopic"  vector subject to the constraint
\begin{equation}
\vec S^{\,2} =1\,.  
\label{ts}
\eeq
The second term in (\ref{senza}) presents a deformation of the O(3) model. 
Sometimes it is referred to as the Skyrme-Faddeev, or Faddeev-Hopf model,
for a review see \cite{manton}.
The constant $F$ has dimension $[m^2]$ while $\lambda$ is dimensionless.\footnote{Below we will introduce a different parametrization in which $F^2 = \xi/2$ and $\lambda = (\beta -1)/g^2$. Moreover,
$g^2\xi \equiv m_\gamma^2$. The origin of this parametrization will become clear shortly.
}

The 
vacuum corresponds to a constant value of $\vec S$ which we can 
choose as $(\vec S )_{\rm vac} =(0,0,1).$ Due to (\ref{ts}) the target space of the
sigma model at hand is $S_2$.
Finiteness of the soliton energy  implies that the vector $\vec S$ 
must tend to its vacuum value at the spatial infinity,
\begin{equation}
\vec S\to \{0,0,1\}\,\,\,\mbox{at}\,\,\, \left| \vec x \right| \to\infty\,.
\label{vac}
\end{equation}
The boundary condition (\ref{vac})
compactifies the space to $S_3$. Since $$\pi_3 (S_2) = {Z}\,,$$ 
the knot solitons present topologically nontrivial maps of $S_3\to S_2$.
As was noted in \cite{Fadd}, there is an associated integer topological
charge, the Hopf invariant, which presents the soliton number. This
charge cannot be the degree of mapping $S_3\to S_2$ because dimensions
of $S_3$ and $S_2$ are different. 

Faddeev and Niemi conjectured \cite{Fadd} that the Hopf solitons can carry a 
twisted toroidal structure and can form knotted configurations. 
Later this was confirmed in numerical studies \cite{batt,N1,N2}. The physical meaning of the Hopf invariant can be visualized in a rather transparent way if we have a large size string-like toroidal structure with windings. Then
the Hopf number should reduce to the topological number in the perpendicular slice times
the number of windings, as shown 
 in Fig.~\ref{donu}
\begin{figure}[th]
\begin{center}
\leavevmode \epsfxsize 8 cm \epsffile{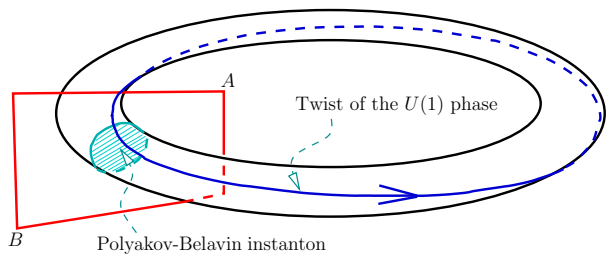}
\end{center}
\caption{{\protect\footnotesize The simplest Hopf soliton, in the adiabatic
limit, corresponds to a Belavin-Polyakov ``instanton" extended in one extra dimension
and bent into a torus, with   a $
2\protect\pi$ twist of the instanton  phase modulus.}}
\label{donu}
\end{figure}
illustrating that the topological stability is enforced due to   a twist of a U(1) phase associated 
with the Belavin-Polyakov instanton
(for a review see e.g. \cite{nsvzrev}).
In other words, the Hopfion picture presumably becomes rather simple  in the limit when the ratio
of the periods is a large number. 

The Hopf solitons  were identified\,\footnote{We will comment more on Hopfions in solid state physics at the end of Sect. \ref{hopfion}.}
  in solid state physics \cite{babaev,babaev2} and in QCD with
quarks in the adjoint representation \cite{bolo}. It was argued \cite{ward}
that an interpolation between a baryon number-2 Skyrmion 
and a Hopf soliton can be found.  For a review
on this subject see \cite{volkov}.
Hopf-type solitons were
discussed in \cite{qtorus} where the  \ntwo U(1) gauge theory (SQED)
was considered.
A theory with several discrete distinct vacua was  engineered   \cite{qtorus} 
supporting domain walls. Then one folds such a  wall into a cylindrical
structure and bends the cylinder to form a torus.
It was argued that the structure thus obtained
is a Hopf-type
soliton 
on the perturbed Higgs branch of the moduli space. The 
soliton was stabilized by a  twist inducing
an Abelian charge. A resurgence of interest to Hopfions is reflected
in the recent publications \cite{N1,N2}. 

In this paper we address some aspects of the Hopf solitons and the Skyrme-Faddeev model. We start with the proof of the following statement: The Skyrme-Faddeev model is the low-energy limit of scalar QED with a potential of a certain type
(inspired by supersymmetric QED). Then we present  arguments (valid provided 
 provided two key parameters are large and based on the Skyrme-Faddeev model {\em per se}
and the underlying parent theory, scalar QED) that the Hopfions of the twisted-toroidal type
do exist.  We also briefly discuss similar constructions
in 4D cylinder geometry.

The paper is organized as follows. In Sec. \ref{preli}
we outline a general picture behind the emergence of the twisted-toroidal solitons
and their relation with the Hopf invariant. In Sec. \ref{bm} we present our basic model: scalar QED with 
two charged flavors. Although this model is not supersymmetric, the form of the scalar field
interacting potential is prompted by supersymmetry. The U(1) gauge group is Higgsed, and in the low-energy limit 
($m_\gamma\to\infty$) we demonstrate the emergence of the Skyrme-Faddeev model
for the massless fields. Thus, the suggested renormalizable model can be viewed as an ultraviolet completion of the
Skyrme-Faddeev model. Section \ref{semistring} is devoted to peculiarities of the ``straight" strings in this model.
In Sec. \ref{hopfion} we explain how to construct a Hopfion of the twisted-toroidal type
by introducing windings and why our analytical consideration is applicable. The applicability requires
choosing two free parameters to be large.
In Sec. \ref{ontc} we fractionalize the Belavin-Polyakov instanton by compactifying one spacial dimension onto $S_1$.
This automatically fixes the size modulus in terms of the size of $S_1$.

\section{Preliminaries}
\label{preli}
\setcounter{equation}{0}

First, let us note that if a Hopf soliton is found in the model with the energy functional
\beq
{\mathcal E} =\int d^3 x\left[ \frac{F^2}{2} \partial_{i}\vec{S}
\,\partial_{i}\vec{S} +\frac{1}{4}\,
\left(\partial_{i} \vec{S}\times\partial_{j}\vec{S} \right)
 \left(\partial_{i}\vec{S} \times\partial_{j}\vec{S} \right)\right]
 \label{21}
\eeq
one can homogeneously inflate all spacial dimensions 
by  passing to
\beq
{\mathcal E} = \int d^3 x\, \sqrt{\lambda}\, \left[ \frac{F^2}{2} \partial_{i}\vec{S}
\,\partial_{i}\vec{S} +\frac{\lambda}{4}\,
\left(\partial_{i} \vec{S}\times\partial_{j}\vec{S} \right)
 \left(\partial_{i}\vec{S} \times\partial_{j}\vec{S} \right)\right]
 \label{22}
\eeq
and then performing the transformations
\beq
x \to \sqrt \lambda \, \,x\,.
\label{23}
\eeq

The Hopf invariant cannot be written as an integral of a local density
-- local in the field $\vec S$. However, if one uses a U(1)
gauged formulation of the CP(1) sigma model in terms of the doublet
fields $n^i,\,\, \bar n_i $ ($i = 1,2$) and
\begin{equation}
\bar n_i\, n^i =1\,, \qquad \vec S = \bar n \, \vec \tau\, n\,,
\label{home}
\end{equation}
(for a review see \cite{nsvzrev,uch}), then the Hopf invariant
reduces to the Chern--Simons term for the above gauge field \cite{Fadd},
\beq
\mathcal{H} = \frac{1}{4\pi ^{2}}\int d^{3}x\,\epsilon ^{\mu \nu \rho }
\left( A_{\mu}\partial _{\nu }A_{\rho }\right), \qquad
A_\mu = -\frac{i}{2}\, \bar n \,\overset{\leftrightarrow}{\partial}_\mu\, n\,.
\label{tue-two}
\eeq

The Hopf solitons that were found numerically \cite{batt} have an intricate knot-like shape. At the same time
the Hopf invariant seemingly has a transparent meaning in the Skyrme-Faddeev model.
Namely, for the torus-like configurations (Fig. \ref{donu})
it reduces to the instanton number (or, alternatively,  magnetic flux) in the perpendicular slice times a winding along the torus large cycle.

To illustrate this interpretation
consider a four-dimensional gauge theory which can be obtained from Witten's superconducting string model
\cite{wit1} by its reduction. Namely, let us downgrade one of two U(1)'s of the Witten model 
to a global symmetry, rather 
than local,
\beqn
{\mathcal L} &=&
 -\frac{1}{4g^2}\,F_{\mu\nu}F^{\mu\nu} +|\cde_\mu\phi |^2-\frac{\lambda_\phi}{4}\,
\left(\phi^2-v_\phi^2
\right)^2
\nonumber\\[2mm]
&+&
|\p_\mu\chi |^2
-\frac{\lambda_\chi}{4}\,
\left(\chi^2-v_\chi^2
\right)^2
-\beta\phi^2\,\chi^2\,.
\label{3}
\eeqn
This model is a crossbreed between those used in \cite{Sh1,Sh2}. 
If the constants $\lambda_{\phi,\chi}$ and $\beta$ are appropriately chosen, the field $\phi $
condenses 
in the vacuum, Higgsing the gauge U(1) symmetry and, simultaneously, stabilizing the field $\chi$. 
Then in the vacuum $\chi_{\rm vac}=0$ which implies that the global U(1)
associated with the $\chi$ phase rotations remains unbroken.
The theory (\ref{3}) obviously
 supports a string which is almost the ANO string. There is an important distinction, however.
 In the string core $\phi =0$, and the $\beta\phi^2\,\chi^2$ term stabilizing $\chi$
 is switched off. Having $\chi=0$ inside the string is energetically inexpedient. Thus, the string solution has
 $\chi \neq0$ in the core \cite{wit1}. This spontaneously breaks the global U(1) on any given string solution.
As a result, a massless phase field $\in {\rm U}(1)$ is localized on the string. The world-sheet theory 
becomes
\beq
S =\int dt\,dz \, \left\{  \frac{T}{2}\left[ (\p_\mu x_0 )^2 + (\p_\mu y_0 )^2\right] + f^2 (\p_\mu\alpha )^2
\right\}
\label{4}
\eeq
where $T$ is the string tension, $f$ is a (dimensionless) constant which can be expressed
in terms of the bulk parameters, $t$ is time, $z$ is the coordinate along the string
while $x_0$ and $y_0$ are perpendicular coordinates. They can be combined as $x_\perp =\{x_1,\,x_2\}$,
where $x_\perp $ depends on $t$ and $z$, 
$$x_\perp = x_\perp (t,z)\,.$$
Moreover, $\alpha(t,z)$ is the phase field on the world sheet, 
$\alpha \leftrightarrow \alpha \pm 2\pi \leftrightarrow \alpha \pm 4\pi \,...\,$. In other words, the target space of $\alpha$ is the unit circle.

Now, let us take a long Abrikosov string and bend it into a circle of circumference $L$.
If $\alpha$ is constant along $z$ (say, $\alpha = 0$), this configuration is obviously 
unstable. Minimizing its energy, the torus will shrink until $L$ becomes of the order of 
the string thickness
$\ell$, and then the string will annihilate. However, one can stabilize it by forcing 
$\alpha$ to wind along $z$ in such a way as to make the full $2\pi$ winding when $z$ changes from 0
to $L$,
\beq
\alpha (t,z) = 2\pi z/L\,.
\label{twist}
\eeq
Note that $\alpha$ linearly depending on $z$ goes through the equation of motion on the world sheet, 
$\p^2 \alpha =0$. For $k$ windings 
\beq
\alpha_k (t,z) = 2\pi z \,k /L\,,\qquad k=2,3,...
\label{ktwist}
\eeq
It is not difficult to estimate the value of $L$. Indeed,
the string energy is\,\footnote{For an alternative idea on obtaining a similar $L+\frac1L$ formula see 
\cite{ni}. }
\beq
E = TL + \frac{(2\pi \, f)^2}{L}
\label{6}
\eeq
Minimizing (\ref{6}) with respect to $L$ we get
\beq
L = 2\pi f/\sqrt{T}\,.
\eeq
Making $f$ large enough we can always force $L$ to be much larger than the flux tube thickness
$\ell$ which is roughly speaking of the order of $1/\sqrt{T}$. Alternatively, we can make $k$ large enough.  For $k$ windings
$L = 2\pi f\, k/\sqrt{T}\,$.

The soliton of the type discussed above was first constructed
 in \cite{Sh1} where it goes under a special name ``vorton'' (in the context of cosmic strings;
for a recent review and a rather extended list of references see
\cite{radu}). Needless to say, in the problem of vortons we do not
have a topological Hopf invariant in the strict mathematical sense of this word. However, the very 
existence of the Hopf-like 
solitons demonstrated above shows that, 
 perhaps, something like a ``quasi"-invariant does exist. Indeed, consider a class
of field configurations in which the field $\chi$ vanishes nowhere except, perhaps, spatial 
infinity.\,\footnote{This is a dynamical requirement, of course,
demanding the toric flux tube's length to be
much larger than its thickness.}
For such field configurations one can define
$\alpha = {\rm Arg}(\chi)$ at all spatial points except, perhaps, infinity.
Then consider the following integral\,\footnote{Note that in this simple example the Hopf-like charge (\ref{gw}) is nothing
but the integral of the charge density component of the conventional 
Goldstone-Wilczek anomalous  current. It can be identified as the anomalous contribution
to the electric or axial charges depending on the parity of the scalar field.
}
\beq
h= \int d^3 x  \,\,c\, \epsilon^{ijk} F_{ij} \partial_k \alpha\,,
\label{gw}
\eeq
where $c$ is a normalizing constant. It is obvious that
$h$  is an integral over a  full derivative.
For field configurations with no zeros of $\chi$ it is well defined. Moreover, if 
for a given field configuration $h\neq 0$ then this field configuration describes, say, a ``twisted torus"
similar to the Hopf solitons. The above twisted torus
 is classically stable. Instability occurs through tunneling.

As was mentioned, the model considered above does {\em not} possess honest-to-god Hopf invariant. Below we will demonstrate 
that a slightly more complicated renormalizable model -- four-dimensional QED, with two scalar flavors and a potential of a special form in the Higgs regime -- 
reduces exactly to the Skyrme-Faddeev model (\ref{senza}) in the low-energy limit $m_\gamma^2\to\infty$, and thus
supports the Hopf invariant. We then formulate a condition under which  a stable vorton exists in this model. 
Although the condition of existence is likely to be met, the corresponding arguments are heuristic rather than rigorous.
The relation between
the vorton in two-flavor scalar QED  and Hopfions obtained numerically e.g. in \cite{batt,N1,N2} (see also
\cite{manton}) remains unclear.

\section{The basic model and its low-energy limit}
\label{bm}
\setcounter{equation}{0}

\subsection{The basic model}

To begin with we will analyze four-dimensional scalar QED with two flavors and the self-interaction potential of a special form. The model is non-supersymmetric but the form of the potential is supersymmetry-inspired.

The  action can be written as follows: 
\begin{equation}
\label{scqed}
S_{\rm 0}=\int d^4x\left\{-\frac1{4g^2}\,F^2_{\mu\nu}+\left |{\mathcal D}_\mu
\vp^A\right|^2- \lambda\left(|\vp^A|^2-\xi \right)^2\right\},
\end{equation}
where ${\mathcal D}_\mu$ is the covariant derivative
\beq
{\mathcal D}_\mu =\partial_\mu - iA_\mu \,,
\eeq
$A$ is the flavor index, $A=1,\,2$, and $\xi$ is a real positive parameter (which can be identified as the Fayet-Iliopoulos term   \cite{FI} in supersymmetric QED). Moreover, for what follows we will introduce
a parameter $\beta$ for the ratio of the coupling constants,
\beq
\beta = \frac{2\lambda}{g^2}\,.
\eeq
In supersymmetric QED we would have $\beta =1$. However, as we will see below, to stabilize the Hopfion in  the semiclassical regime one must require $\beta\gg1$.
The vacuum energy in (\ref{scqed}) vanishes. The vacuum manifold is determined by
\beq
\left|\vp^1\right|^2 + \left|\vp^2\right|^2 = \xi\,.
\label{215}
\eeq
The above constraint leaves us with three real parameters out of four residing in $\vp^{1,2}$. 
One extra (phase) parameter can be eliminated by imposing an appropriate gauge condition. It is easy to see that the vacuum manifold is nothing other than $S_2$, presenting the target space of the CP(1) model. The U(1) gauge boson is Higgsed. The spectrum of the model consists of a massive photon, a massive Higgs meson,
\beq
m_\gamma = \sqrt{2} g\sqrt{\xi}\,, \qquad m_H = m_\gamma \,\sqrt{\beta}\,,
\eeq
and two massless Goldstone particle corresponding to oscillations on the vacuum manifold. Below we will be interested in the limit $\beta \gg 1$ or, alternatively, $m_H\gg m_\gamma$, known as the London (or Abrikosov) limit in the theory of the ANO strings.

The model (\ref{215}) supports semilocal strings (see e.g. \cite{sy,SY-BOOK}). Their core is provided by the 
Abrikosov-Nielsen-Olesen string \cite{ANO}, while the tail is due to the Belavin-Polyakov two-dimensional instanton 
\cite {BP} of the
CP(1) model lifted in four dimensions.

\subsection{Low-energy limit}

If excitation energies are lower than $m_\gamma$, the photon and the Higgs boson
can be integrated out. Then we obtain the low-energy theory for the moduli fields which, as was mentioned, reduces
to the CP(1) sigma model.

In fact, at $\beta=1$ (i.e. in the supersymmetric limit) the answer is known. If we introduce  normalized $n$ fields,
 \beq
 n^A = \frac{\vp^a}{\sqrt \xi}\,,\qquad \bar n_A n^A =1\,
 \eeq
 in the gauged formulation the low-energy model for the moduli fields is the standard CP(1) model,
 \beq
 {\mathcal L} = \xi \left({\mathcal D}_\mu \bar n_A\right)^\dagger \left({\mathcal D}^\mu   n^A\right)\,.
 \eeq
 The field $A_\mu$ in the definition of the covariant derivative is auxiliary; 
 it has no kinetic term and is expressible in terms of $n^A$
(see e.g. Sec. 27.2 in \cite{uch}),
\beq
A_\mu = -\frac{i}{2} \left(\bar n_A\stackrel{\leftrightarrow}\partial_\mu n^A
\right).
\label{219}
\eeq
The relation between $S^a$ in Eq. (\ref{senza})  and $n$ is
\beq
S^a = \bar n\tau^a  n\,,\qquad a= 1,2,3,
\label{220}
\eeq
where $\tau^a$ are the Pauli matrices. 

Now we will show that at $\beta\neq 1$ an additional term is generated in the low-energy action -- the Skyrme-Faddeev term in (\ref{senza}). 

The simplest way to establish the existence of the Skyrme-Faddeev term is to evaluate the two actions (microscopic and its low-energy limit) in a string background field satisfying the Bogomol'nyi equations:

In the microscopic theory (\ref{scqed}) these equations are
\beq
F_{12} +g^2 \left(|\vp^A|^2 -\xi
\right)= 0\,,\qquad \left( {\mathcal  D_1} +i {\mathcal D_2}
\right)\vp^{1,2} =0\,.
\eeq
The string is assumed to be aligned along the $z$ axis.
If these equations are satisfied, the microscopic action takes the form
\beq
 \Delta S =  \int d^4x \, \frac{g^2}{2} (\beta -1) \left(|\vp^A|^2-\xi \right)^2 = \int d^4x  \frac{\beta -1}{2g^2}\,  F_{12}^2 \,,
\label{222}
\eeq
This expression can be obviously uplifted to four dimensions
\beq
\Delta S = \int d^4x  \frac{\beta -1}{4g^2}\,  F_{\mu\nu}^2\,.
\label{223}
\eeq
Generally speaking, Eq. (\ref{222}) could  miss possible four-derivative terms that vanish on the Bogomol'nyi (anti)selfdual fields.
One can see, however, that none of such terms can be uplifted to four dimensions in the
 Lorentz invariant way. Equation (\ref{223}) is the only exception.  Therefore
 the result (\ref{223}) is complete.
 
 Thus, we conclude that the low-energy theory for the scalar QED (\ref{scqed}) is
 \beq
 S = \int d^4x \left\{ \frac{\xi}{4}\pt_{\mu}S^a \pt^{\mu}S^a - \frac{\beta -1}{4g^2}\,  F_{\mu\nu}F^{\mu\nu} +... \right\}\,,
 \label{leaction}
 \eeq
 where $F_{\mu\nu}$ should be expressed in terms of $S^a$ using (\ref{219}), (\ref{220}), namely,
 \beq
 F_{\mu\nu} = \frac12\,\varepsilon_{abc}S^a \pt_{\mu}S^b \pt_{\nu}S^c\,.
 \label{fmunu}
 \eeq
 Then
 \beq
 F_{\mu\nu}^2 = \frac{1}{4}\left( \varepsilon_{abc} \partial_\mu S^b \partial_\nu S^c\right)^2\,,
 \eeq
 cf. Eq. (\ref{senza}).
The ellipses in  Eq. (\ref{leaction})  stand for higher derivative corrections (with six derivatives and higher). The theory in (\ref{leaction})
 is the Skyrme-Faddeev model (\ref{senza}), with a particular choice of parameters.

\section{Semilocal Abelian strings}
\label{semistring}
\setcounter{equation}{0}

In  scalar QED  with two flavors (see Eq.  (\ref{scqed})) strings are no longer the conventional 
ANO strings with exponentially small tails of the
profile functions. The presence of massless fields in the bulk makes them semilocal.
The semilocal  strings have a power fall-off at large distances from the string axis (see  \cite{AchVas} for a review). The semilocal BPS string
interpolates between the ANO string and two-dimensional  O(3) sigma-model instanton
uplifted to four dimensions (also known as the lump). The semilocal string possesses
an additional zero mode associated with string's transverse size $\rho$.
In the limit  $\rho\to 0$ we recover the ANO string while
at $\rho \gg 1/m_{\gamma}$ it becomes a lump.

\subsection{Semilocal string solution}

Consider first the BPS-saturated semilocal string in the theory with $\beta=1$.
The ansatz for  the string solution  has the following structure:
\begin{eqnarray}
\label{profsl}
\vp^1(x) &=& \phi_1 (r)\, e^{i\,\theta}\ ,\nonumber\\[2mm]
\vp^2(x) &=& \phi_2 (r)\, ,\nonumber\\[2mm]
A_i(x) &=&\! \!-\varepsilon_{ij}\,\frac{x_j}{r^2}\ [1-f(r)]\,,
\\[2mm]
r&\equiv &\sqrt{\vec x_\perp^{\,2}}\,,\nonumber
\end{eqnarray}
with the  boundary conditions implying that only one scalar field
has a nonvanishing condensate inside the core
\begin{eqnarray}
\label{boundary}
&&\phi_1 (0)=0 \,, \qquad f(0)=1\ ,\nonumber\\[2mm]
&&\phi_2 (\infty)=0\,, \qquad
\phi_1(\infty) =\sqrt{\xi}, \qquad f(\infty)=0 \,,
\end{eqnarray}
Here $r$ and $\theta$ are polar coordinates in the plane orthogonal to the string axis.

It is not difficult to find an approximate solution valid at large values of the scale modulus,
\beqn
\phi_1(r)
&=&
\sqrt{\xi}\frac{r}{\sqrt{r^2+|\rho|^2}}\, ,
\nonumber\\[2mm]
\phi_2(r)
&=&
\sqrt{\xi}\frac{\rho}{\sqrt{r^2+|\rho|^2}}\,,
\nonumber\\[2mm]
\label{lump}
f
&=&
\frac{|\rho|^2}{r^2+|\rho|^2}\,.
\eeqn
with the complex modulus $\rho$. The absolute value $|\rho |$ is the scale modulus, while 
arg$\,\rho$ is a phase modulus inherent to the Belavin-Polyakov lump.
This solution  is related to  the Belavin-Polyakov instanton solution in the two-dimensional O(3) sigma model 
uplifted to four dimensions.

\subsection{Effective world-sheet theory}

Substituting
the static solution in  the original four-dimensional action (\ref{scqed})
and assuming that the modulus $\rho$ has a slow adiabatic
dependence on the world-sheet coordinates $t$ and $z$ we get the following answer:
\beq
\ce_{N_f=2}\, \, =2\pi\,\xi\int dt\,dz\,\, 
|\pt_k\rho|^2
\,\,\ln{\frac{L}{|\rho|}}\,,
\label{abel}
\eeq
where $k=0,3$ labels the world-sheet coordinates. This action has
only the kinetic term and no potential term. This is because $\rho$
is associated with the exact zero mode of the string solution
for the BPS string.
 Note, that the integral over $r$ is logarithmically divergent in the infrared; a cut off is provided by
 the string by the length $L$ (which is supposed to be very large but finite).
 We  work  in the logarithmic approximation and assume that
 $L\gg |\rho|$. A similar effective low-energy theory was obtained in \cite{sy} for non-Abelian semilocal strings.
 
 Recently logarithmic divergence of the norm of the size zero modes for semilocal vortex in (2+1) dimensions was addressed in
 \cite{IntSeib13}. It was shown that in (2+1) dimensions there is a superselection rule for vortices with
 different $\rho$. In our (3+1) case the vortex become a string, and its finite length $L$ provides a physical infrared
 cut-off for the logarithmic divergence. The logarithmic factor in (\ref{abel}) is large but still finite.
 
In what follows we will  consider a
more generic case with $\beta >1$, and in particular, the large-$\beta$ limit. 
If $\beta \neq 1$ the string  becomes non-BPS: it  tends to shrink in type I superconductors 
($\beta<1$)  and   expand at $\beta>1$, which corresponds to type-II superconductivity. 

To calculate the part of the world-sheet energy functional  due to violation
of ``BPS-ness" we use the low-energy action (\ref{leaction}) taking into account the four-derivative correction.
Substituting the gauge potential (\ref{profsl}), (\ref{lump}) into the second term in (\ref{leaction}) 
we arrive at   the following
effective  world-sheet theory on the semilocal string 
\beq
\ce_{\rm eff}= 2\pi\,\int dt\,dz\left\{
\xi\,|\pt_k\rho|^2
\,\,\ln{\frac{L}{|\rho|}}+
\frac13\,\frac{(\beta -1)}{g^2}\, \frac{1}{|\rho|^2}\,\right\}.
\label{2add}
\eeq
Note that this $\rho$ dependence  emerges from the lower limit 
of  integration of the profile function over the
radial coordinate transverse to the string. Since the instanton profile function presented above
is invalid at small $\rho$ Eq.~({\ref{2add})  
cannot be trusted at $\rho\rightarrow 0$.  In fact we 
need
\beq
|\rho|\gg \frac1{m_{\gamma}}
\label{largerho}
\eeq
to justify the last term in (\ref{2add}).

The second term in the action above shows that the  string
is unstable. The  thickness $\rho$ of the semilocal string in the type-II superconductor ($\beta>1$) expands.
Note, that this term was obtained  in \cite{hind} in the limit $(\beta -1)\ll 1$ where the  stability issue was first discussed. 
The formula (\ref{2add}) is valid for all $\beta$ as long as $|\rho |^2 \gg (\beta -1)m_\gamma^{-2}$. 
In Sec. \ref{hopfion} we will show that in fact the $|\rho |^{-2}$  term in (\ref{2add}) is the first term of expansion
of a function of which some general features can be established.

\section{Making a Hopfion}
\label{hopfion}
\setcounter{equation}{0}

Previously, the topologically stable solutions in the model (\ref{senza}) saturating the Hopf invariant 
were found by numerical calculations. Here we would like to outline some analytic argument regarding the  stability of a ``large" Hopfion of the twisted-toroidal type.

As was explained in Sec. \ref{semistring}, a semilocal string solution of scalar QED (see (\ref{scqed})) has a 
a complex modulus $\rho \equiv |\rho| e^{i\alpha}.$ The size $|\rho |$ is arbitrary.
Now, we bend this string in the form of a torus, and let the phase $\alpha$ wind along the  torus as shown in Fig.~\ref{donu}. More exactly, this figure shows one winding while in fact it can be any integer number, to be denoted by $k$.
To justify the analytic consideration below we need
to stabilize $|\rho |$ at a large value such that $|\rho|\gg m_\gamma^{-1}$, and, in addition, to stabilize the
circumference $L$ at $L\gg|\rho|$. There is an interplay of various factors to be analyzed. 

First, the string tension $T=2\pi\xi$ produces a term linear in $L$ in the twisted-toroidal soliton mass, $\delta M_1 = 2\pi\xi L$ . Second, the winding
\beq
\alpha (z)= \frac{2\pi k z}{L}
\label{twisti}
\eeq
generates 
the term
\beq
\delta M_2 = 2\pi \xi\,\int_0^L dz \,|\rho|^2 \left(\frac{\partial\alpha}{\partial z}\right)^2\, \ln \frac{L}{|\rho|} = \frac{\xi(2 \pi|\rho | k)^2}{L}\,\ln \frac{L}{|\rho|}.
\eeq
If $\delta M_1$ tends to shrink $L$, the second term $\delta M_2$ tends to make $L$ larger, provided that the value of 
$|\rho|$ is fixed. The problem is that it is not fixed for the time being. 

The question of the $|\rho|$ stabilization depends on the $|\rho |^{-2}$ term in (\ref{2add}). Now we will assume
 $\beta \neq 1$ and analyze  its impact. It turns out that at $\beta \sim 1$ the size $|\rho|$ is stabilized at $|\rho|\sim 1/m_\gamma$ where our approximation is invalid. In order to achieve stabilization at larger scales 
 we must assume that $\beta \gg 1$. 
 
 However, at large $\beta$ the second
term in (\ref{2add}) is no longer a small correction to the first one. At large $\beta$ we need to sum up all
higher derivative corrections enhanced by powers of $\beta$. It is not difficult to see that
calculating higher order terms, were such a calculation possible, would produce a series of the type
\beq
\delta M_3 = \frac{\beta\,\xi L}{3m_\gamma^2|\rho|^2}\, \sum_{\ell = 0}^\infty \, c_\ell \left(\frac{\beta}{|\rho|^2 m_\gamma^2}
\right)^\ell\equiv  \frac{\beta  \,\xi  L}{3m_\gamma^2|\rho|^2}\, F\left(\frac{\beta}{|\rho|^2 m_\gamma^2}
\right)
\label{329}
\eeq
where $F$ is some function of its argument 
\beq
\kappa =  \frac{\beta}{|\rho|^2 m_\gamma^2}\,,
\label{kappa}
\eeq 
and we dropped unity compared to $\beta$ in $\beta -1$. What is known about the function $F$?

At $\kappa\to 0$ the function $F\to 1$. One can can argue that $F$ stays finite for all values of $\kappa$. 
Moreover, using the scaling analysis it is easy to see that if stabilization is possible at all it should occur at
$\kappa \sim 1$. Finally, we expect that at large $\kappa$
\beq
F(\kappa) \sim \frac32\, \frac{\ln{\beta}}{\kappa}, \qquad \kappa \to\infty
\label{55}
\eeq
to match the Abrikosov formula for the string tension in the London limit. Indeed, then 
at large $\kappa$ (small $\rho$) the tension of the untwisted string will
approach the Abrikosov formula $T = 2\pi \xi \log\beta $ for the tension of the ANO string in the limit $\beta\gg 1$ \cite{ANO}. This tension  is much larger than the first term $2\pi\xi$ ( tension of the ANO BPS string) because of the $\log\beta$ enhancement.

Strictly speaking, we cannot descend down to $\kappa\lsim 1$ with Eq. (\ref{55}). However, qualitatively, 
it seems safe to say that $F(\kappa\sim 1)\sim \log\beta\gg 1$ being enhanced by  $\log\beta$. Since $F(0)=1$ one can conclude then that $F$ increases as $\kappa$ varies from zero to unity, so that $F^\prime $ is parametrically larger
(on average)\footnote{The prime in $F^\prime $ above denotes
differentiation with respect to $\kappa$.} than unity at $\kappa\lsim 1$. Once $\kappa$ becomes $\gsim 1$ the function $F$ starts decreasing, as dictated by (\ref{55}). The expected qualitative behavior of the function
$F(\kappa)$ is shown in Fig.~\ref{functionF}.

\begin{figure}[th]
\begin{center}
\leavevmode \epsfxsize 6 cm \epsffile{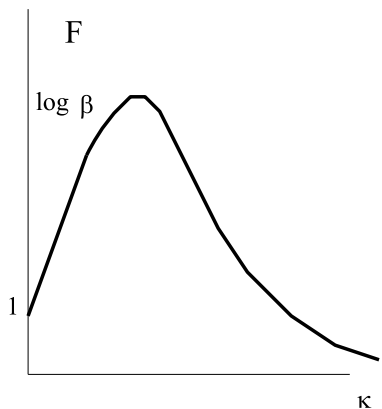}
\end{center}
\caption{{\protect\footnotesize Qualitative shape of the function $F(\kappa)$.}}
\label{functionF}
\end{figure}

Assembling all three terms together
we get for
 the energy of the twisted semilocal string
\beq
{\mathcal E} = 2\pi\,\left\{
\xi L + \frac{\xi(2 \pi|\rho | k)^2}{L} \,\, \ln \frac{L}{|\rho|}+ \frac{\beta }{3g^2}\,
\frac{L}{|\rho|^2}\,F\left(\frac{\beta}{|\rho|^2 m_\gamma^2}\right)\right\}.
\label{etl}
\eeq
The system of the extremization equations with respect to $L$ and $\rho$ is
\begin{eqnarray}
&&
1 - 4\pi^2 k^2\,\frac{|\rho |^2 }{L^2}\,\ln \frac{L}{|\rho |}  + \frac{2\beta}{3 m_{\gamma}^2|\rho|^2} \,F(\kappa)= 0 \,,\nonumber\\[2mm]
&&
4\pi^2 k^2\,\frac{|\rho |^2 }{L^2}\,\ln \frac{L}{|\rho |}  - \frac{2\beta }{3 m_{\gamma}^2|\rho|^2}\,F(\kappa) -
\frac{2\beta^2}{3 m_{\gamma}^4|\rho|^4}\,F'(\kappa)= 0 \,,
\label{rhoLeqs}
\end{eqnarray}
where we assume that the logarithmic factor $\log L/|\rho |$ is large and need not be differentiated.
Adding these two equations we find
\beq
\kappa^2\,F'(\kappa)= \frac32.
\label{kappaeq}
\eeq 
Provided the solution of this equation is  found, the value of $L$ is stabilized at
\beq
\frac{L^2}{|\rho|^2}= \frac{4\pi^2 k^2\,\ln{k}}{1+\frac23\kappa F(\kappa)}.
\label{L}
\eeq
The energy of the Hopfion is 
\beq
{\mathcal E} = \frac{4\sqrt{2}\pi^2\,\sqrt{\beta}}{g}\,\sqrt{\xi}\,
k\,\sqrt{\log k}\,\sqrt{\frac{1+\frac23\kappa F(\kappa)}{\kappa}}.
\label{hopfenergymin}
\eeq

The argument presented above tells us that  equation (\ref{kappaeq}) has two solutions, one at $\kappa_{*} \sim 1/\sqrt{\log\beta}$ and another one at $\kappa_{**} \sim 1$. The qualitative behavior of left and right hand sides of the equation
\beq
F'(\kappa)= \frac32\,\frac1{\kappa^2}
\label{kappaeq1}
\eeq
are shown in Fig.~\ref{fig:kappaeq}.

\begin{figure}[th]
\begin{center}
\leavevmode \epsfxsize 6 cm \epsffile{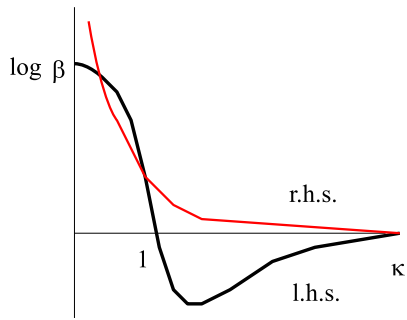}
\end{center}
\caption{{\protect\footnotesize Left and right hand sides of Eq (\ref{kappaeq1}).}}
\label{fig:kappaeq}
\end{figure}

It is easy to see that the second solution corresponds to the maximum of  energy,
while the minimum  is achieved at
\beq
\kappa_{*} \sim \frac{1}{\sqrt{\log\beta}}, \qquad
|\rho_* | \sim \frac{\sqrt{\beta}}{m_\gamma}\,(\log\beta)^{1/4}\,, \qquad 
\frac{L_*}{|\rho_*|}\sim k\sqrt{\log k} \,
\eeq
The energy of this solution is 
\beq
{\mathcal E} \sim \frac{\sqrt{\beta}}{g}\,\sqrt{\xi}\,
k\,\sqrt{\log k}\,(\log \beta)^{1/4}.
\label{hopfenergy}
\eeq

The conditions of applicability of the approximations we made are met provided $k \gg 1$ and $ \beta\gg 1$.

This concludes our consideration demonstrating the existence of the twisted-toroidal Hopfions with $L\gg\rho$ in the 
scalar QED (\ref{scqed}).

Before passing to some extensions we would like to pause here to make a comment on  related studies 
carried out in the context of condensed matter physics. An earlier discussion relevant to our consideration 
carried out in the two-flavor scalar QED (see (\ref{scqed}))  can be found in an important paper  
 \cite{babaev2}. This paper demonstrates, in particular,  that  two-flavor scalar QED is equivalent  to the Faddeev-Skyrme model
(\ref{senza}) coupled to a massive vector field.
A  perturbative procedure was suggested in    \cite{babaev2} which allows one to integrate out
the above vector field. This  leads to certain higher derivative corrections to the
Faddeev-Skyrme model (six derivatives or higher). The result for higher derivatives in \cite{babaev2} matches
 our Eq. (\ref{leaction}) obtained in a totally different way. This shows that
the Faddeev-Skyrme model is  a low-energy limit of two-flavor
scalar QED. 

Moreover, it was also argued in \cite{babaev2} that a large Hopfion could exist as a
local minimum in the full theory (\ref{scqed}), while in the Faddeev-Skyrme
model {\em per se} the Hopfion soliton is a topological soliton and represents
a global minimum. The latter statement also matches  our analysis demonstrating  the
presence
of the Hopfion solution at $L\gg\rho\gg 1/m_{\gamma}$.

Details of our analysis and that of \cite{babaev2} are quite different. In particular, in \cite{babaev2}
it is the kinetic term that plays a crucial role, while our focus is on both the kinetic and potential terms. We plan to
elaborate on this elsewhere.
The important step in our Hopfion solution analysis  is the
construction of the
two-dimensional effective theory (\ref{2add})  on the semilocal string.

\section{Replacing \boldmath{$R_4$} by $R_3\times S_1$:  ``composite''   \\[1mm] 
twisted semilocal strings}
\label{ontc}
\setcounter{equation}{0}

As was discussed above, the  semilocal string under discussion can be viewed
as an interpolation between the local ANO  string in the core
and the two-dimensional Belavin-Polyakov instanton uplifted in four dimensions. 
By compactifying one of spatial dimension into $S_1$
one can split the Belavin-Polyakov CP(1) instanton in two ``constituents" 
which, in turn, could 
provide one with possibilities for  
additional winding numbers for such closed strings. 

Let us recall a well-known fact: four-dimensional BPST instanton melts
upon compactification of one of coordinates \cite{const} (see also \cite{unsal,Dunne}):
it dissociates into constituents with fractional topological charge $1/N$ (two constituents
with the topological charge 1/2 in SU(2)).
 The constituents 
carry the  instanton number $1/N$ as well  as a monopole number.
The gauge field holonomy  is generated at one-loop level;
 the eigenvalues of the holonomy fix the distance between the constituents
along the compact direction.  As a result, the instanton in the SU$(N)$ gauge theory 
with one compactified dimension (the so-called caloron, see \cite{vanb})
turns out to be a composite object built from   $N-1$ ``conventional'' monopoles
(with the $1/N$ instanton charge) and the so called Kaluza--Klein (KK) monopole \cite{const}. The latter 
``wraps around" the compact dimension.

A similar type of behavior for two-dimensional instantons in CP(1)
was discussed too \cite{mnit,bruk} (see also \cite{mnit1}).  In particular, if  the spatial dimension is compactified, 
then the CP$(N-1)$ sigma-model instanton (generalizing the Belavin-Polyakov instanton) splits into $N$ constituents. 
The scale instanton modulus   acquires a 
new interpretation: it represents the distance between two constituents
(for CP(1)), while the sizes of the constituents
are fixed by the radius of the compact dimension. 

Now, while uplifting the instanton from two to four dimensions, let us make an intermediate  stop at three dimensions.
  The two-dimensional CP$(N-1)$ instantons are the baby Skyrmion
particles in three dimensions . 
 Recently it was recognized \cite{t3} (see also \cite{Eto:2009bz})
that these baby Skyrmions have a transparent composite structure: they 
consist of $N$ ``ultraviolet" degrees of freedom which, naively, had been 
integrated out. The picture is 
 similar to that of the baryon in the four-dimensional 
chiral Lagrangian (the Skyrme model) \cite{sky}.  A composite structure of the Skyrmion
was argued  in a number of complementary ways which did not necessarily include  the compact
dimension.

The important question is: what are the quantum numbers of  the ``partons"  
 discussed in \cite{t3}?  It was argued that, just like in other similar considerations,
the constituents \cite{t3} are ordered along the roots of SU$(N)$. In addition,  there is the KK constituent
which corresponds to the affine root. These constituents have the topological charges
$Q=1/N$ with respect to $\pi_2$,  and   nontrivial charges with respect to the
magnetic Abelian group. The presence of the additional charges can be most
easily seen upon dualizing the BPS equation \cite{t3}.  To this
end,  it is convenient to dualize one scalar field through the Polyakov relation \cite{Polya}
\beq
\partial_{\nu}\phi = \epsilon_{\nu\mu\rho}F_{\mu\rho}\,.
\label{1duo}
\eeq
Taking into account the explicit form of the  instanton solution
we obtain an analog of the Gauss law,
\beq
d^{*}F = (\delta(z-z_{+}) - \delta(z-z_{-})) \,.
\label{gl}
\eeq
where $z_\pm$ stand for the position of two constituents.

Thus,   two constituents of the  CP(1) instanton with the
opposite charges are clearly seen in the dual formulation. These constituents in the
dual formulation are point-like.
The total charge of the Skyrmion with respect to the dual photon  vanishes.
This explains the origin of the permanent binding 
of two constituents inside the Skyrmion. Indeed, the ``electric" energy of each parton 
diverges logarithmically (in 2+1 dimensions),
while the energy of the electrically neutral Skyrmion compound is finite. 

With this knowledge in hand, we can make the final uplift from three to four dimensions.
Since we start from  the composite Skyrmion in $D=3$, we will
finish with the composite semilocal string in $D=4$. This composite strings
should have $N$ partonic ``substrings," with fractional  fluxes $1/N$.
This 
corresponds to
uplifted topological charges, as well as additional ``flavor"
fluxes. There is also a special  KK substring corresponding to the 
uplifting of the KK Skyrmion.  The total ``flavor flux" of the semilocal
string vanishes while the  total magnetic flux $Q=1$.

For instance, the composite semilocal string in uplifted CP(1)  involves one
fractional substring while the second substring is of the Kaluza-Klein type. The topological charges
of each of the two constituents are 1/2. Hence, the total tension of the composite string 
comes from two equal parts. It is convenient to write down the instanton solution
in the following form 
\cite{belavin}:
\beq
\omega= \frac{z-z_{+}}{z-z_{-}}\,,
\label{love}
\eeq
where the moduli $z_{+}$ and $z_{-}$ correspond to the positions of the partons.
The center of mass and the scale moduli are defined as
\beq
Z = \frac{1}{2}\left(z_{+} + z_{-}\right)\,,\qquad  \rho = |\rho | e^{i\theta}= \frac{1}{2}\left(z_{+} - z_{-}\right)\,.
\label{bela}
\eeq
It is clearly seen that the scale modulus represents the distance between
the partons. To reveal the additional quantum numbers of these two
constituents it is convenient to use an analog of the trick with 
dualization of the BPS equation  \cite{t3}. In four dimensions
the scalar field is dual to a rank-two field $B_{\mu\nu}$,
\beq
\partial_{\rho} \phi = \epsilon_{\rho \alpha \mu \nu} \partial_{\alpha}B _{\mu \nu}\,.
\label{ranktwo}
\eeq
As a result,  from the BPS equation we immediately get
\beq
d^{*}H = (\delta(z-z_{+}) - \delta(z-z_{-}))\,,
\label{imme}
\eeq
where $H_{\alpha\mu\nu}= \partial_{\alpha}B _{\mu \nu}\,.$
Equation (\ref{imme})  clearly shows that the two constituent substrings have the opposite
charges with respect to the $B$ field. 

Next, we fold the composite semilocal string into a toroidal structure 
with the simultaneous  twist of the 
complex scale modulus.
Since the   modulus $|\rho |$ is now interpreted as the distance between the two constituents
its nontrivial twisting corresponds a kind of  linking of the constituents. 
This ``linking number" plays the role of the topological number responsible for the
classical stability of the soliton.\footnote{
Note that one could also imagine twist with a $Z_N$ rotation since $Z_N$ does not 
act on the instanton solution in the CP$(N-1)$ model.}

To conclude this section note that the instanton splitting   naturally happens
for the twisted-toroidal string provided we compactify one spatial dimension.
The Belavin-Polyakov instanton splits
into composites with fractional topological charges. The 
closed strings  under discussion are twisted, and the twist amounts
to a nontrivial holonomy along the compact dimension.

\section{Conclusions}
\label{conclu}

We demonstrated that the Faddeev-Skyrme model emerges as a low-energy limit
of scalar QED with two charged scalar fields and a selfinteraction potential
of a special form (inspired by supersymmetric QCD). 
Our conclusion parallels that previously made in the condensed matter literature \cite{babaev2},
although both, motivations and derivations, are different.
Then we discuss
possible Hopf solitons of the``twisted-toroidal" type. We need to stabilize both the 
size of the Belavin-Polyakov instanton (appearing in the perpendicular slice) and the length $L$.
We presented analytical arguments that such stabilization is achieved provided
$\beta \gg 1$ under the condition that the number of windings is large two. 

Then we briefly discussed a similar twisted toroidal construction in four dimensions with one spatial dimension
compactified.

\section*{Acknowledgments}

We are grateful to E. Babayev, A. Niemi, M. Nitta, and Ya. Shnir for useful remarks.

The work of M.S. is supported in part by DOE grant DE-FG02- 94ER-40823. 
The work of A.Y. is  supported 
by  FTPI, University of Minnesota, 
by RFBR Grant No. 13-02-00042a 
and by Russian State Grant for 
Scientific Schools RSGSS-657512010.2.
A.G. thanks FTPI at the University of Minnesota 
where a part of the work was carried out, for  hospitality and support.
The work  of A.G. was supported in part by the Grants No.
RFBR-12-02-00284 and PICS-12-02-91052.

\end{document}